# Realignment of ferroelectric nematic by photoinduced electric field

- Short title: **Realignment of ferroelectric nematic by light**

Sathyanarayana Paladugu[1], Oleksandr Kurochkin[2,3], Ruslan Kravchuk[2,3], Mykola Kravets[2,3], Volodymyr Sashuk[2,3], Bijaya Basnet[1,4], Vassili G. Nazarenko[2,3]*, Sergij V. Shiyanovskii[1], and Oleg D. Lavrentovich[1,4,5]*

[1] Advanced Materials and Liquid Crystal Institute, Kent State University; Kent, OH 44242, USA.
[2] Institute of Physics, National Academy of Sciences of Ukraine; Kyiv, 03028, Ukraine.
[3] Institute of Physical Chemistry, Polish Academy of Sciences; Warsaw, 01-224, Poland.
[4] Materials Science Graduate Program, Kent State University; Kent, OH 44242, USA.
[5] Department of Physics, Kent State University; Kent, OH 44242, USA.

*Corresponding Authors. Emails: vnazaren@iop.kiev.ua, olavrent@kent.edu

**Abstract**

Liquid crystal displays (LCDs) employ thin pixels of a paraelectric nematic, confined between two glass plates with transparent indium tin oxide (ITO) electrodes. A voltage applied to the electrodes realigns the molecules, changing the optical appearance of the pixel. Here we demonstrate that a similar molecular realignment can be triggered by low-power light irradiation when the paraelectric nematic in the cell with ITO electrodes is replaced with a ferroelectric nematic ($N_F$). Irradiation creates an intrinsic electric field that is strong enough to realign the $N_F$ but too weak to affect the paraelectric nematic. The extraordinary sensitivity of the ITO/$N_F$ pair to light irradiation creates a platform for optical control of ferroelectric polarization and applications in photonics, sensing, smart windows, and beyond.

**Teaser**

Light irradiation of a ferroelectric nematic cell with indium tin oxide electrodes realigns the vector of electric polarization.

## INTRODUCTION

Rod-like molecules of a conventional paraelectric nematic (N) align parallel to each other, along a common direction, called the director, $\hat{\mathbf{n}}$. The director is an apolar unit vector, $\hat{\mathbf{n}} \equiv -\hat{\mathbf{n}}$, since the N order does not distinguish between the heads and tails of the molecules even when these are chemically different (*1*). Because of the orientational order, the properties of the N, such as dielectric permittivity, measured along $\hat{\mathbf{n}}$ and in perpendicular directions, are different. This dielectric anisotropy is at the heart of electro-optics employing the N phase. A thin N slab is confined between two glass plates coated with ITO electrodes and alignment layers that align $\hat{\mathbf{n}}$ along a predesigned in-plane direction. An electric field applied to the electrodes across the slab creates a torque proportional to the dielectric anisotropy and the square of the field, which realigns $\hat{\mathbf{n}}$, changing the cell appearance thanks to birefringence of the N. This realignment of $\hat{\mathbf{n}}$, called the Fréedericksz effect, requires fields on the order of $10^6$ V m$^{-1}$.

The recently discovered ferroelectric nematic ($N_F$) exhibits a polar orientational order that sets a spontaneous polarization **P** collinear with $\hat{\mathbf{n}}$ (*2-5*). An external electric field aligns **P** parallel to itself. The realigning torque is proportional to the field amplitude and is thus stronger than its N counterpart: **P** realigns by the fields on the order of $10^2$ V m$^{-1}$ (*4, 6*).

In this work, we demonstrate realignment of the $N_F$ polarization in a cell with ITO electrodes irradiated by low-power low-intensity visible or ultraviolet (UV) light beams. Irradiation of ITO creates a weak (tens of V/m) in-plane electric field that realigns **P**. The light-induced realignment of polarization in the presence of ITO expands dramatically the richness of physical effects in the $N_F$ phase and holds a major promise for applications.

## RESULTS

We use an $N_F$ mixture FNLC919 (*7-12*) with the following phase sequence on cooling from the isotropic (I) phase: I – 82 °C – N – 46 °C − SmZ$_A$− 32 °C − $N_F$ – 8 °C – Crystal, where SmZ$_A$ stands for an antiferroelectric smectic phase. Similar results were obtained for an $N_F$ material DIO (*3*), with the phase sequence (*13*) I - 174 °C - N - 82 °C - SmZ$_A$ - 66 °C – $N_F$ - 34 °C - Crystal. The data are presented mostly for FNLC919, since its room-temperature range of the $N_F$ phase is easier to characterize; exploration of the temperature gradient effect is more convenient for the higher-temperature DIO. The $N_F$ layer is confined between two glass plates with ITO electrodes and rubbed polyimide PI2555 alignment layers. The rubbing direction $\mathbf{R} = (0, -1, 0)$ of PI2555 sets a uniform monodomain orientation of polarization $\mathbf{P}_0 = P(0,1,0)$, in the $xy$ plane of the cell, Fig.1A (*8, 14*). The thickness $d$ of $N_F$ layer is in the range $(3 - 6)$ µm. A higher thickness might result in spontaneous twist caused by depolarization effects described recently (*15-18*). A monochromatic visible or UV light beam normally impingent on the cell realigns **P** , Fig. 1, as detailed below.

### Light-induced realignment of polarization in planar $N_F$ cells

Irradiation of the $N_F$ cell with ITO electrodes realigns **P** away from $\mathbf{P}_0$, forming two lobes that transmit probing light through the cell viewed between two crossed polarizers $\mathcal{P}$ and $\mathcal{A}$ of a polarizing optical microscope (POM), Fig.1B,C. The qualitative character of reorientation does not depend on beam polarization: beams of linear polarization $\mathcal{L}$ perpendicular, Fig.1B,D and parallel, Fig.1C,E to **P** yield similar results. Regardless of the angle between $\mathcal{L}$ and **R**, the two lobes are divided along the rubbing direction **R**. In some cells, the lobes are asymmetric, which is probably caused by a slight misalignment of the rubbing directions at the top and bottom plates of the cell. The dynamics of the transmitted light intensity caused by the realigned **P** is exponential, with characteristic times on the order of seconds, Fig.1D-F. Irradiation of a cell with only one ITO electrode produces a weaker effect, Fig.1D,E. A higher beam power, Fig. 1F, and a shorter wavelength $\lambda$ of irradiation cause a stronger effect, Fig.1G-J. The geometry of the two lobes depends on $\lambda$ and power. To characterize this dependence, we introduce the azimuthal angle $\theta$ calculated from $\mathbf{P}_0$, Fig.1H. Irradiation with a longer wavelength, $\lambda = 514$, 488 nm, creates two separate lobes of realignment, with a preserved alignment $\mathbf{P}_0$ along $\theta = 0, 180°$, Fig.1G,H. However, irradiation with a shorter wavelength $\lambda$ =476, 457 nm, but of the same power of 5 mW, enlarges the lobes so that they touch along a zigzag defect at $\theta = 180°$, Fig.1I,J. A similar zigzag defect forms when a power increases for a fixed $\lambda$. At $\theta = 0$, **P** is always along $\mathbf{P}_0$, Fig.1I,J.

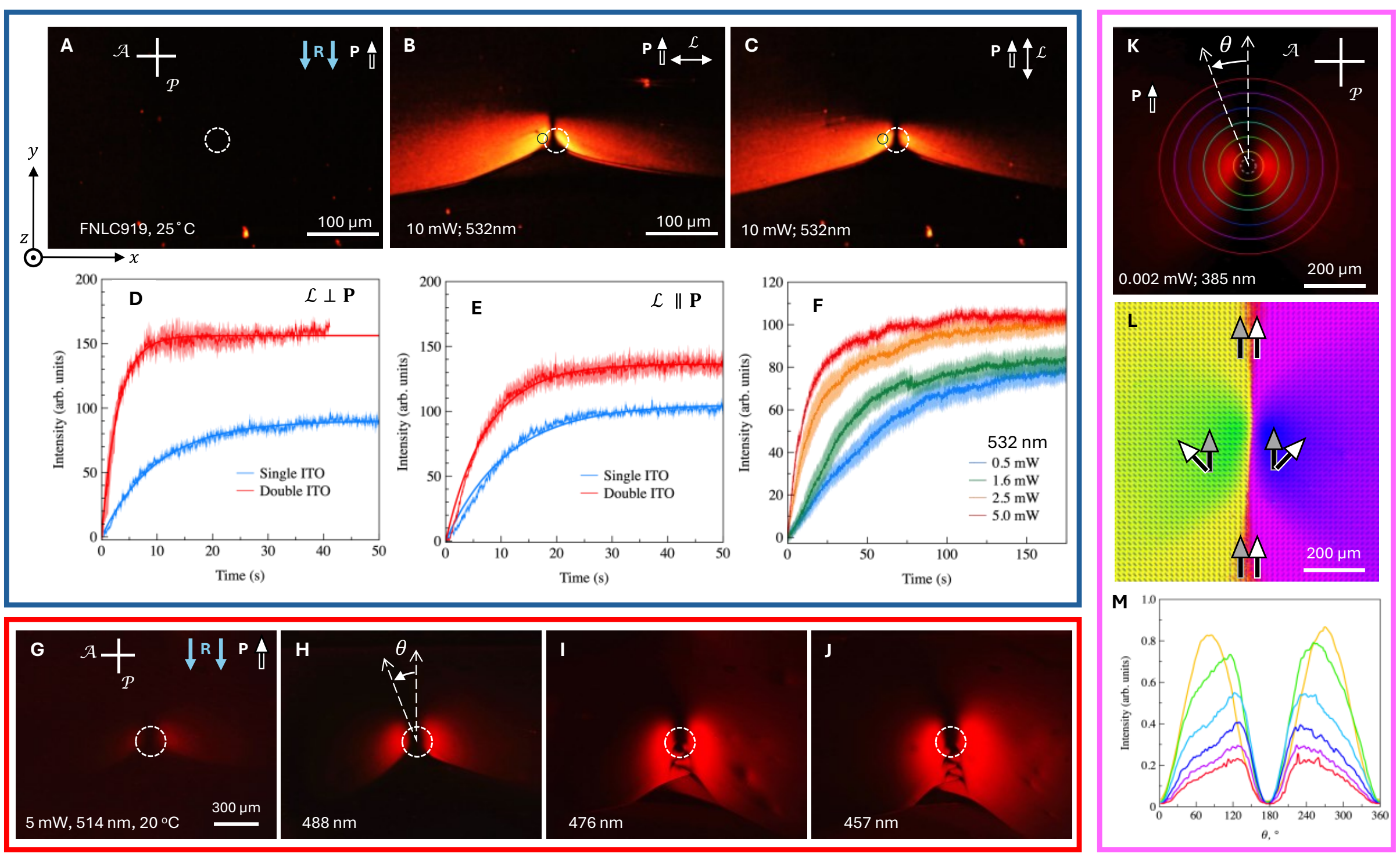

**Fig. 1. Light-induced realignment of P in planar FNLC919 $\mathbf{N_F}$ cells with ITO electrode**s. (**A**) POM texture of the initial planar texture; crossed polarizer $\mathcal{P}$ and analyzer $\mathcal{A}$. (**B,C**) Realignment of **P** caused by 532 nm, 10 mW irradiation with linear polarization $\mathcal{L}$ orthogonal to the rubbing direction and parallel to it, respectively. (**D,E**) Transmitted light intensity through an $\mathrm{N_F}$ cells vs. time for two directions of $\mathcal{L}$ measured within the black circles in parts (B,C) and for cells with one and two ITO electrodes; solid lines: exponential fits. (**F**) Time-dependent transmitted light intensities as a function of laser power for 532 nm irradiation with one ITO electrode. (**G,H,I,J**) Wavelength dependence of realignment patterns caused by irradiations with λ = 514, 488, 476, and 457 nm, respectively. (**K**) POM texture of realignment by low-power 2 μW UV irradiation, 385 nm. (**L**) PolScope texture of the same cell and the same realigned pattern as in (K) showing reorientation of the optic axis; white arrows show the realigned **P**, grey arrows show $\mathbf{P}_0$. (**M**) Transmitted light intensity $T$ vs. the azimuthal angle $\theta$ measured in the texture (K) along the colored circles; **P** turns counterclockwise in the left lobe and clockwise in the right lobe. $d$ =5 μm; 25 °C in parts (A-F); $d$ = 4.5 μm; 20 °C in parts (G-M). White dashed circles mark the beam waist, 36 μm in parts (A-C), 200 μm in parts (G-J) and 40 μm in (K). PI2555 alignment layers.

The light intensity $I$ needed to realign **P** is low, for example, $I = 0.2\ \mathrm{W\ mm^{-2}}$ for $\lambda = 457$ nm in Fig.1J. The UV effect is even stronger: at $\lambda$ =385 nm, realignment is well-developed at $I = 1.8 \times 10^{-3}$ W mm$^{-2}$, Fig.1K-M. The spectral efficiency correlates with the enhancement of light absorption by ITO at shorter wavelength: a 100 nm thick ITO layer absorbs 5% at $\lambda$ =500 nm and 10% at 400 nm (*19*). The realignment effect is also enhanced by lowering the temperature of sample, which is explained by increase of the polarization amplitude $P$, Fig.S1.

The presence of both the polar order and ITO electrodes is the necessary condition of the light-induced realignment of **P**. Irradiation of $N_F$ cells without ITO does not realign **P**. There is no realignment when the material is in the antiferroelectric $SmZ_A$ or the paraelectric N phase. The principal mode of polarization realignment in the planar cells is an azimuthal deviation from the initial orientation $\mathbf{P}_0$, as illustrated by the PolScope texture in Fig.1L.

The symmetry of polarization pattern in Fig.1L and the Gaussian geometry of the beam suggest that the light-induced realignment of **P** is caused by an axially symmetric agent acting in the $xy$ plane of the cell. There are two candidates associated with light absorption by the ITO, which is a doped n-type semiconductor with a bandgap between 3.5 and 4.4 eV (*20, 21*): a thermal gradient and an electric field **E**. A 100 nm thick ITO layer absorbs a few percent in the spectral range (400-1200) nm (*19*), which is sufficient to heat by tens of degrees (*22*). A generation of an electric field at the interface between an illuminated ITO electrode and adjacent organic materials (*23, 24*), including the paraelectric N (*21, 25*), has been also documented. Irradiation with light in the lower visible and UV parts of spectrum creates electron-hole pairs at the ITO surface with electrons moving away from the illuminated spot (*20, 23*). The efficiency of charge generation and absorption increase at shorter wavelength, which agrees with the dependency in Fig.1G-K. Below, we discuss the relevance of these two factors in the observed light-induced realignment of **P**.

## Light-induced heating

The thermal gradient was reported (*26*) to realign **P** in planar $N_F$ cells with ITO electrodes subject to a 445 nm laser irradiation. However, experiments described below do not support this mechanism as the cause for the patterns created by visible and UV light in Fig.1.

i. Placing the DIO $N_F$ cell at the edge of a heating element, to create a temperature gradient, does not cause realignment, Fig.S2. DIO is chosen for this experiment because of its high temperature range of the $N_F$ phase. The $N_F$ layer of a thickness 3.5 μm is confined between two thin, 150 μm, glass cover slides with the ITO and alignment layers. The horizontal temperature gradient along the direction normal to the edge of hot stage is determined by measuring the location of the border lines separating the $SmZ_A$ from the N and $N_F$ phases at a function of the hot stage temperature, Fig.S2. This temperature gradient is $2 \times 10^4$ °C m$^{-1}$, four times higher than the gradient $0.5 \times 10^4$ °C m$^{-1}$ reported in Ref. (*26*) to realign **P** in $N_F$ cells. In other words, the thermal gradient is not the prime reason for the realignment effect illustrated by Fig.1.

ii. When the ITO is etched to form an ITO-glass interface and this interface is irradiated, the realignment is always at the "south" side of the spot, regardless of whether this side is glass or ITO, Fig.S3. If the effect were caused by heating, one would expect the reorientation on either the ITO or the glass side, as these two are of different thermal conductivity, $\approx$ 4 W m$^{-1}$ °C$^{-1}$ for ITO (*27*) and $\approx$ 1 W m$^{-1}$ °C$^{-1}$ for glass (*28*). The patterns support the idea of irradiation-generated in-plane radial electric field **E**. At the north side of patterns in Fig.S3, **E** is directed towards north. Since **P** is directed from south to north, mostly parallel to **E**, the realignment is weak in the north part. In the south part, **P** is mostly antiparallel **E** and the realignment is stronger.

iii. Irradiation of cells with thick, 180 nm, and thin, 25 nm, ITO electrodes produces practically identical patterns, despite the 7-fold difference in the absorption pathway, which is significant as the coefficient of absorption by ITO is $\alpha \approx 5 \times 10^5\ \mathrm{m}^{-1}$ at 532 nm (*29*).

iv. An infrared (IR) laser beam, $\lambda = 1064$ nm, focused to a diameter 70 μm, of a power below 15 mW, does not produce any realignment of **P**, Fig.2A. Power above 20 mW causes substantial heating (*22*) and phase transitions from the $\mathrm{N_F}$ to higher-temperature phases, Fig.2B-E. Textural symmetry of the IR irradiated cell in Fig.2B-E is different from the asymmetric textures triggered by visible and UV irradiation in Fig.1. Namely, in the $\mathrm{N_F}$ region that surrounds the heated spot with the $\mathrm{SmZ_A}$, N, and the isotropic phases, $\mathbf{P}(x, y)$ forms circular arches tangential to the circular $\mathrm{N_F}$ interface, to avoid deposition of surface bound charges. The circular arches are separated from the uniform planar regions by parabolic defects, Fig.2B-F, described previously (*14, 30-33*). Similar heating effects and parabolic defects can be observed when the ITO electrodes are replaced with gold, and irradiated with a $\lambda$ =532 nm laser beam, Fig.2G,H. The distance ~ 50 μm over which the irradiated spot shows coexisting isotropic, N, $\mathrm{SmZ_A}$ and $\mathrm{N_F}$ phases, allows us to estimate the temperature gradient as $\sim 10^5\ ^\circ\mathrm{C\ m^{-1}}$ at 20 mW of IR irradiation in Fig.2B. For a similar 1064 nm 20 mW laser beam of a waist 16 μm, irradiating a 100 nm ITO layer in a similar $\mathrm{N_F}$ cell, Andolšek et al. (*22*) measured the temperature gradients to be as high as $0.7 \times 10^7\ ^\circ\mathrm{C\ m^{-1}}$. These gradients are much higher than the value $0.5 \times 10^4\ ^\circ\mathrm{C\ m^{-1}}$ reported to cause a heating-induced realignment of **P** in Ref. (*26*). Thermal gradients are thus not the prime reason for the observed realignment **P** caused by visible and UV light in Fig.1.

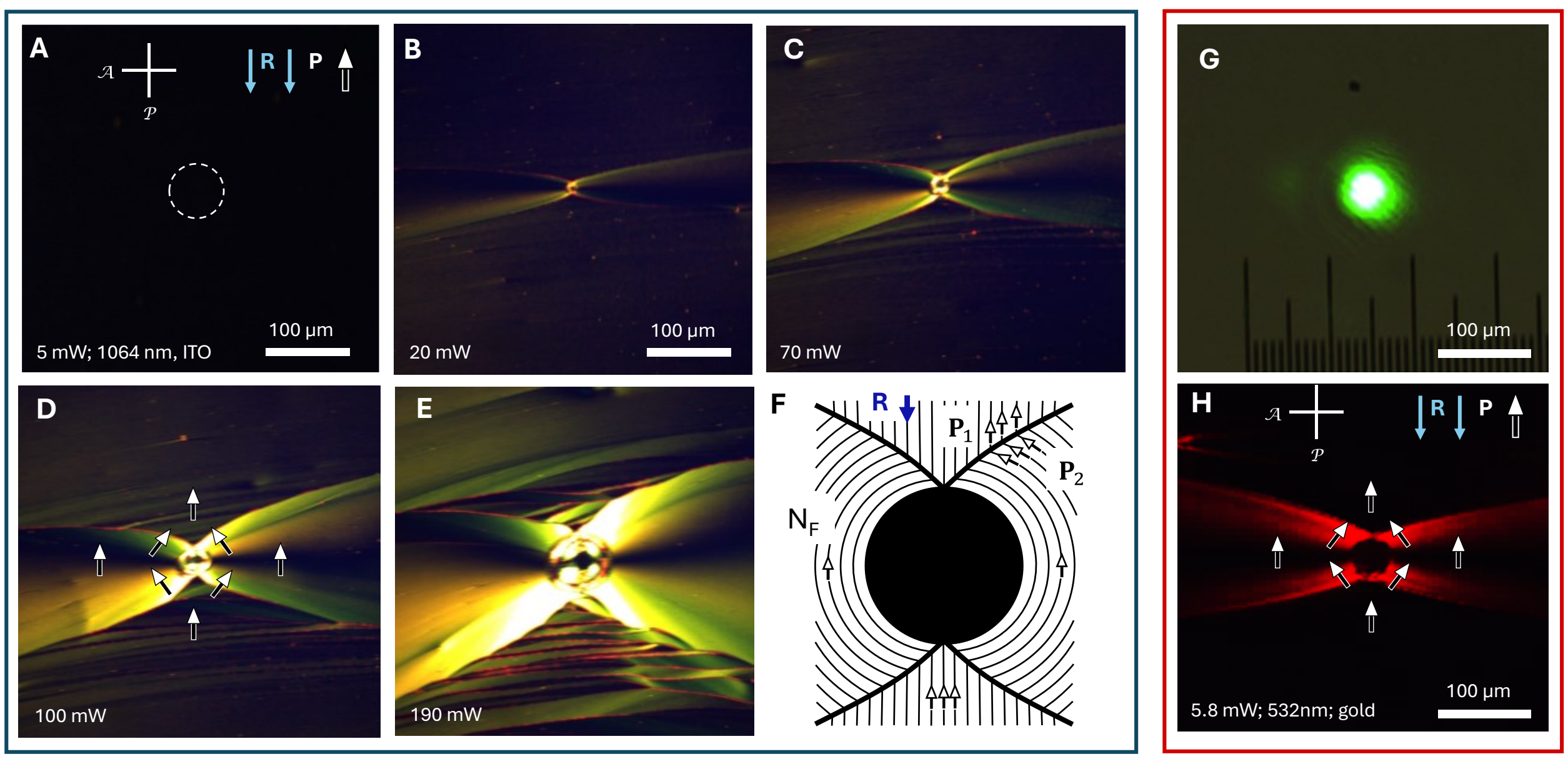

**Fig. 2. Polarization reorientation caused by the light-induced heating.** (**A-E**) Infrared-irradiated region in planar $\mathrm{N_F}$ FNLC919 cells with ITO substrates. $\lambda$ = 1064 nm, diameter 70 μm, with various powers from 5 mW to 190 mW. $d = 4.5$ μm, T = 20 °C. (**F**) Scheme of polarization field in the $\mathrm{N_F}$ cell with a circular inclusion that sets a tangential anchoring of **P**; the uniform polarization field $\mathbf{P_1}$ set by the rubbing direction **R** and the circular polarization field $\mathbf{P_2}$ are separated by domain walls in the shape of parabola branches; the shape is dictated by the avoidance of bound charge at the domain walls. (**G**) 70 μm diameter laser beam, $\lambda$ = 532 nm, power 5.8 mW, irradiating (**H**) an $\mathrm{N_F}$ cell with gold electrodes. PI2555 alignment layers.

v. A weak (500 V $m^{-1}$) in-plane direct current (DC) electric field $\mathbf{E}_{\text{pol}}$ is applied to the cell during the cooling from the N phase over 10-20 min. The field $\mathbf{E}_{\text{pol}}$ is removed before light irradiation. These electrically poled samples show a much higher sensitivity to the visible and UV light than the samples not subjected to $\mathbf{E}_{\text{pol}}$. For example, a poled sample, $d =$ 3.3 µm, realigns by a 532 nm beam of an intensity $I = 6.5 \times 10^{-3}$ W mm$^{-2}$ (power 5 mW, waist 1 mm), while a similar unpoled sample in Fig.1L requires 1 W mm$^{-2}$. When a shorter wavelength, 405 nm, is used, the intensity sufficient to realign **P** in poled samples drops to $5 \times 10^{-4}$ W mm$^{-2}$ (power 0.12 mW, diameter 550 µm), Fig.S4. The explanation is that $\mathbf{E}_{\text{pol}}$ reduced the number of free ions in the sample. Such a reduction should not alter the heating effect but is instrumental if the light beam creates an intrinsic electric field in the ITO.

## Light-induced intrinsic electric field

The experiments suggest that the main cause of the polarization realignment is a radial electric field **E**, tangential to the $xy$ plane, created in the ITO by visible and UV light irradiation (*21, 23-25*). The electric field component $E_z$ normal to the cell is screened by a small tilt of **P** away from the $xy$ plane (*34, 35*). Electric charging caused by irradiation of the $N_F$ planar cell with ITO electrodes is verified by direct measurements of a photocurrent in cells with the alignment layers, Fig.S5, and without them, Fig.S6. The photoinduced electric field in N cells with ITO electrodes has been recently reviewed by Habibpourmoghadam (*21*). For a Gaussian beam, as suggested by Fig.1L, **E** points outward the center of the illuminated spot, $\mathbf{E} = E(r)(\sin\theta, \cos\theta, 0)$, where $\theta$ is the angle between **E** and $\mathbf{P}_0$, and $r$ is the radial distance from the center.

The amplitude $E$ is zero at the center of irradiated spot (the flat tip of the Gaussian peak does not produce a charge gradient), increases with $r$ and tends to zero away from the spot. When the ITO electrode during irradiation is held at a biasing potential in the range $[-2\text{V}, +2\text{V}]$, the pattern of polarization realignment is the same as the one at zero potential.

The in-plane field imposes a realigning torque $\boldsymbol{\tau} = \mathbf{P} \times \mathbf{E}$ on the polarization, which is zero whenever **E** is collinear with **P**, at $\theta = 0,\ 180°$. The sense of rotation of **P** is opposite in the left and right lobes, since the signs of $\boldsymbol{\tau}$ are opposite, Fig.1L. To describe the patterns in Fig.1B,C,G-L, we first find the local azimuthal angle $\varphi(z)$ between the field-realigned **P** and the surface anchoring imposed $\mathbf{P}_0$, from the balance of elastic twist and electrostatic energy per unit area, considering only the $z$-dependences:

$$f = \int_0^d \left\{\frac{1}{2} K_2 \varphi'^2(z) - E(r) P cos\big(\theta - \varphi(z)\big)\right\} dz, \tag{1}$$

where $\varphi' = \frac{d\varphi}{dz}$ and $K_2$ is the twist elastic constant. Variation of (1) results in

$$\varphi'^2(z) = \frac{2E(r)P}{K_2}\left[cos(\theta - \varphi_m) - cos\big(\theta - \varphi(z)\big)\right], \tag{2}$$

with the incomplete elliptic integral of the first kind $F(\varphi, k)$ as a general solution:

$$\sqrt{\tilde{E}}\,\frac{2\lfloor z - z_m\rfloor}{d} = \int_{\varphi}^{\varphi_m} \frac{d\tilde{\varphi}}{\sqrt{cos(\theta - \varphi_m) - cos(\theta - \tilde{\varphi})}} = F\left(\sin^{-1}\sqrt{\frac{1-(u/k)^2}{1-u^2}}, k\right), \tag{3}$$

where $\varphi_m = \varphi(z_m)$ is the maximum twist angle, $u = \cos\frac{\theta-\varphi}{2}$, $k = \cos\frac{\theta-\varphi_m}{2}$, and $\tilde{E} = \frac{E(r)Pd^2}{2K_2}$ is the dimensionless field. Because of the symmetric boundary conditions $\varphi(0) = \varphi(d) = \varphi_b$, $z_m = d/2$, the maximum twist $\varphi_m = \varphi(d/2)$ obeys the equation

$$F\left(\sin^{-1}\sqrt{\frac{1-(b/k)^2}{1-b^2}}, k\right) = \sqrt{\tilde{E}}, \tag{4}$$

where $b = \cos\frac{\theta-\varphi_b}{2}$.

The boundary value $\varphi_b$ is controlled by the surface anchoring and twist elastic torques. The in-plane anchoring potential $W(\varphi)$ for $N_F$ cells exhibits two energy minima, one global at $\varphi = 0$, and a local one at $\varphi = \pm\pi$ (*14, 36*):

$$W(\varphi) = \frac{W_Q}{2}\sin^2\varphi + W_P\,(1-\cos\varphi), \tag{5}$$

where $W_Q, W_P \geq 0$ are the quadrupolar apolar and polar anchoring coefficients, respectively. The balance of anchoring torque (*37*), $\left.\frac{dW(\varphi)}{d\varphi}\right|_{\varphi=\varphi_b} = \sin\varphi_b\,(W_Q\cos\varphi_b + W_P)$, and the elastic $K_2\varphi'$ torque, Eq.(2), provides the condition for the surface angle $\varphi_b$ between **P** and $\mathbf{P}_0$:

$$\sin^2\varphi_b(\widetilde{W}_Q\cos\varphi_b + \widetilde{W}_P)^2 = 4\tilde{E}[\cos(\theta-\varphi_m) - \cos(\theta-\varphi_b)], \tag{6}$$

where $\widetilde{W}_{Q,P} = W_{Q,P}d/K_2$ are the dimensionless coefficients. Equations (4) and (6) define $\varphi_m$ and $\varphi_b$.

Figures 3A-C show the dependencies of $\varphi_b(\theta)$, $\varphi_m(\theta)$, and nonlinearity parameter $k(\theta)$ on the azimuthal direction $\theta$ of **E**, respectively. The plots are calculated from Eqs.(4),(6) with $\widetilde{W}_Q = 9$ and $\widetilde{W}_P = 0.9$ that correspond to the previously established estimates (*14*). There is no realignment at $\theta = 0$, where **E** is parallel to $\mathbf{P}_0$. For $\theta = 180°$, the realignment shows a threshold. For an infinitely strong surface anchoring, the threshold field is $\tilde{E}_F = \pi^2/4$. For $\tilde{E} < \tilde{E}_F$, $\varphi_b(180°) = \varphi_m(180°) = 0$, whereas for $\tilde{E} > \tilde{E}_F$, $\varphi_b(180°), \varphi_m(180°) > 0$ and rising with $\tilde{E}$, Fig.3A,B. Such a Fréedericksz-like behavior at $\theta = 180°$ explains the transition from two disjoined lobes in Fig.1G,H to two lobes in contact separated by a zigzag defect in Fig.1I,J. The zigzag defect is necessitated by the opposite sense of rotation of **P** in the two lobes, Fig.1L. With $\tilde{E}_F = \pi^2/4 \approx 2.5$, $P = 4\times10^{-2}$ C m$^{-1}$, $K_2 = 5$ pN, $d = 5$ μm, one estimates the field causing the zigzag pattern as $E \approx 25$ V m$^{-1}$. There is no threshold for realignment at $0 < |\theta| < 180°$. The field $E \approx 25$ V m$^{-1}$, is rather weak to cause a measurable z-component of **P**. For comparison, an alternating current field of a much higher amplitude 330 V m$^{-1}$ applied across a planar cell, could induce only a minor tilt of **P**, by about 0.2° away from the $xy$ plane in the bulk of cell (*6*).

The lack of south-north symmetry of the optical patterns is enhanced by the behavior of the nonlinearity parameter $k = \cos\frac{\theta-\varphi_m}{2}$ in Fig.3C. For strong fields, $k$ is close to 1 and does not depend much on $\theta$. It means that **P** realigns along **E** and $\varphi \approx \varphi_m$ everywhere along the $z$-axis, except for two thin subsurface layers, where $\varphi \to \varphi_b$. For weak fields, $k$ is close to 1 only for small $\theta$. In the south part, $k$ is small, which means that **P** twists smoothly $\varphi \approx \varphi_b + (\varphi_m - \varphi_b)\sin\frac{\pi z}{d}$, thus diminishing transmission of light since two sequential smooth twists of opposite handedness along the $z$-axis restore the linear polarization of transmitted light. Note also that the Mauguin number, $Mau = p\Delta n/\lambda$ is noticeably larger than 1, which helps to preserve the light polarization; here $\Delta n = 0.22$ is the birefringence of the $N_F$ phase measured at $\lambda = 633$ nm, and $p$ is the pitch of twist, estimated to be ~10 µm for $\varphi_m = 90°$ at $d/2 = 2.5$ µm, which yields $Mau = 3.5$.

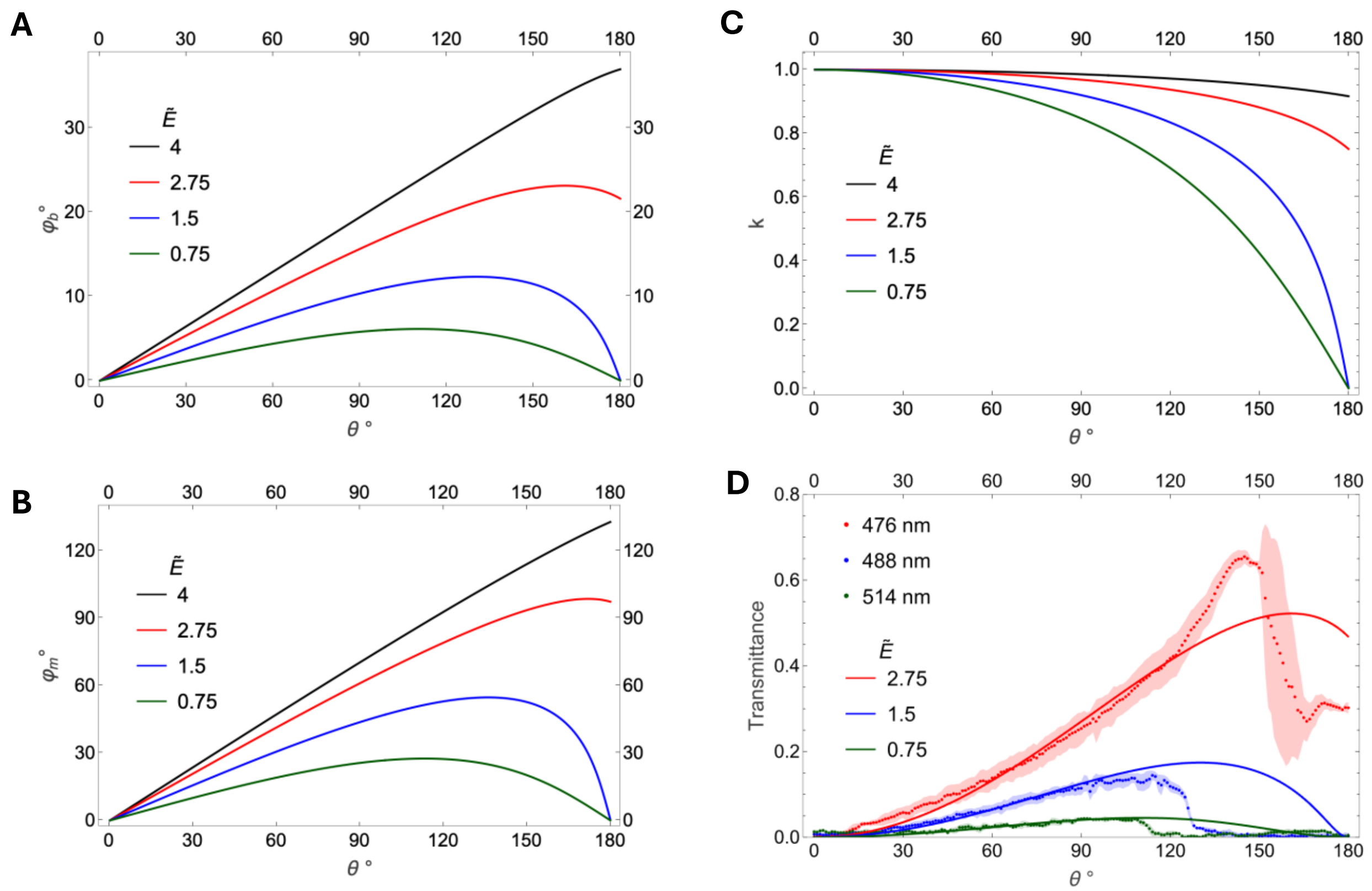

**Fig. 3**. **Polarization twists as a function of the polar angle $\theta$.** Dependences of (**A**) $\varphi_m$ , (**B**) $\varphi_m$, and (**C**) parameter $k$ on the electric field direction $\theta$ for different amplitudes $\tilde{E}$. (**D**) $\theta$-dependence of light transmission for the experimental patterns in Fig.S7, measured along a circular arch with coordinates $\theta = (0, 180°)$, $r = (220 - 260)$ µm for irradiation at 476, 488, and 514 nm by a beam of diameter 200 µm, power 5 mW and $\theta$-dependence of light transmission calculated using the data in part (B) and Eq. (7).

The light intensity transmitted through the pattern and two crossed polarizers, as a function of polar angle $\theta$, is determined mostly by the surface rotation $\varphi_b(\theta)$ of polarization, as (*37*)

$$T(\theta) = \sin^2 2\varphi_b(\theta). \quad (7)$$

Comparison of the experimental and the numerical patterns in Fig.3D shows a good agreement for $0 < \theta < 110°$, thus supporting the model. The discrepancies seen in the south part, $120° < \theta < 180°$, are caused by the following factors. First, the model does not consider deformations in the $xy$ plane. This omission is not important for the north part of the pattern where the realignment is weak but worsens the prediction for the south part where $\varphi_m$ and $\varphi_b$ reach their maximum and the east and west lobes of opposite rotation handedness interact elastically. In particular, the model does not describe the complex deformations caused by the zigzag defect that produces strong oscillations of transmission in the range $150° < \theta < 180°$ in Fig.3D. Second, the model operates with a simple anchoring potential that might not capture the actual anchoring properties when $\varphi_b$ is not small, which is the case of the south part, Fig.3A.

## DISCUSSION

To conclude, we uncovered a strong light-induced realignment of electric polarization in a ferroelectric nematic cell with thin ITO layers. Irradiation of the ITO by a low-power visible or UV light beam creates an intrinsic radial in-plane electric field that realigns the polarization. In planar cells, an unidirectionally aligned polarization experiences clockwise and counterclockwise rotations away from the "easy axis" of alignment.

The sensitivity of the $N_F$ cells to light irradiation is much higher than the sensitivity of a paraelectric N. The N shows no response to the irradiation under the reported conditions but can realign either because of the so-called optical Fréedericksz transition (OFT) or photoelectric interface activation (PEIA) (*38-41*).

An OFT occurs when a laser beam of a high intensity $I_{Fr} \sim 10\ \mathrm{W\,mm^{-2}}$ irradiates a transparent N (*39, 40*); the required intensity drops to $\sim 1\ \mathrm{W\,mm^{-2}}$ for an N material doped with absorbing dyes (*41*). The OFT typically occurs in very thick samples, $d \sim 100\ \mu\mathrm{m}$, since $I_{Fr} \propto 1/d^2$ (*40*). For thin cells, $d = 5\ \mu\mathrm{m}$, the intensities are one-two orders of magnitude higher. In our case, the realignment of a $d = 4.5\ \mu\mathrm{m}$ $N_F$ cell occurs at much lower intensities: for example, $I = 1.8 \times 10^{-3}\ \mathrm{W\,mm^{-2}}$ in Fig.1K.

A PEIA refers to an N version of a photorefractive effect, in which an N cell with electrodes is subject to a static electric field and a light irradiation that mitigates ionic screening of the field, thus causing a realignment of N (*21, 25, 42-44*). The intensities needed for modification of the electric properties can be low, $7 \times 10^{-4}\ \mathrm{W\,mm^{-2}}$ (*42*) and $10^{-5}\ \mathrm{W\,mm^{-2}}$ (*43*), but the realignment requires the presence of the static field and thus the associated conductive leads, batteries, generators, etc.

A relatively slow (seconds) realignment of polarization in the $N_F$ phase should be addressed in future studies with a focus on material properties of the photosensitive component and surface anchoring at the $N_F$-photosensitive interface.

Light-realigned polarization pattern in the form of spiral vortices was recently discovered by Lucchetti et al. (*45*) for $N_F$ slabs confined between two properly cut 1 mm solid ferroelectric crystals of iron-doped lithium niobate that produce a photovoltaic field under intensities (0.5-5) $\mathrm{W\,mm^{-2}}$. The photoinduction of electric field phenomenon is expected to be similar in the two cases, but the described set-up with thin ITO electrodes offers advantages such as compact design and the possibility to uniformly align the $N_F$ and control the surface anchoring.

Realignment of liquid crystals by electric fields is an important phenomenon widely used in applications ranging from the flat-panel LCDs to generation of optical vortices. In the described $N_F$ phenomenon, the realignment is local, as the electric field is created within the illuminated spot, and does not require electronic drivers, wires, external generators, etc. The strong sensitivity of the $N_F$ electric polarization to light, simplicity of the experimental set-up, possibility to design the polarization patterns by tailoring the geometry of ITO coatings, surface anchoring, and light beams promise potential applications in all-optical devices, photonics and electro-optics.

## MATERIALS AND METHODS

Materials. The room temperature $N_F$ mixture FNLC919 was supplied by Merck Electronics KGaA, Darmstadt Germany. The $N_F$ material DIO (*3*) was synthesized in the laboratory following the scheme in Ref. (*46*).

Substrate preparation. We use glass substrates with ITO electrodes of thickness 100 nm and sheet resistivity 10 ohm $sq^{-1}$. The plates are ultrasonically cleaned in distilled water and isopropyl alcohol, dried at 95 ℃, cooled to room temperature, blown with nitrogen, and coated with alignment layers.

The aligning agent PI2555 and its solvent T9039, both purchased from HD MicroSystems, are combined in a 1:9 ratio. The solution is spin-coated onto cleaned ITO glasses according to the following scheme: 1 s @ 500 rpm → 30 s @ 1500 rpm → 1 s @ 50 rpm. After spin coating, the sample is baked at 95°C for 5 min, followed by 60 min at 275°C. The spin coating produces a PI2555 alignment layer of a thickness 50 nm.

The PI2555 layer is buffed unidirectionally using a Rayon YA-19-R rubbing cloth (Yoshikawa Chemical Company, Ltd, Japan) of a thickness 1.8 mm and filament density $280/mm^2$ to achieve a homogeneous planar alignment. An aluminum brick of a length 25.5 cm, width 10.4 cm, height 1.8 cm and weight 1.3 kg, covered with the rubbing cloth, imposes a pressure 490 Pa at a substrate and is moved ten times at a speed of 5 cm $s^{-1}$. The rubbing length is about 1 m. This treatment aligns the director $\hat{\mathbf{n}}$ of a paraelectric N parallel to the rubbing direction $\mathbf{R}$ with a small pretilt $\approx 3° \pm 1°$. In the Cartesian coordinates, the $y$-axis is antiparallel to $\mathbf{R}$, and the $z$-axis is normal to the cell. In the $N_F$ phase of both DIO and FNLC919, the rubbing direction $\mathbf{R} = (0, -1, 0)$ aligns the polarization antiparallel to itself, $\mathbf{P}_0 = P(0,1,0)$, as verified by applying an in-plane electric field and switching the direction of $\mathbf{P}$. These experiments also suggest that the pretilt of $\mathbf{P}_0$ at the PI2555 substrates is negligibly small since the optical retardance does not change when the field is applied.

Cell assembly. Planar cells are constructed from two PI2555-covered glass substrates, with parallel rubbing directions $\mathbf{R} = (0, -1, 0)$. The cell gap thickness $d$ is set by glass spherical spacers in the range (3-6) μm, as measured at five different locations by interferometry using a spectrometer Lambda 18 (Perkin Elmer). The standard deviation of $d$ is less than 0.1 μm. The ferroelectric material is filled into cells along $\mathbf{R}$ in the isotropic phase by the capillary action. The temperature is controlled by a Linkam hot stage with an accuracy of $\pm 0.1$°C.

Experimental set-up. Figure 4 illustrates an experimental set-up based on an POM. Monochromatic laser light L1 of the wavelength 457, 476, 488, 514 nm (tunable laser, Ion Laser Technology), 532nm (Coherent Verdi-V6), and 1064 nm (Hi Q laser) irradiates the cells. The laser power is measured before entering the cell using a laser power meter (PM100D, Thorlabs). The incident intensity of the laser is controlled by placing a circular variable neutral density (ND) filter between the laser and polarizer P1. Polychromatic light sources L2 and L3 serve to observe the textures in the transmitted and reflected modes, respectively. The laser beam impingent onto the cell is of a diameter varied from 36 μm, Fig.4B,C, Supporting Information, to 1 mm using objectives with different magnification. A Thorlabs complementary metal-oxide semiconductor (CMOS) camera records optical textures. The POM observations were performed using a long-pass filter with a cutoff at 575nm (Thorlabs). For 1064 nm, an IR blocking filter (TF1, Thorlabs) is used.

UV irradiation is produced by integrating the PRO6500S (Wintech of Texas Instruments) with the DLP6500 chip, equipped with a 385nm LED light, in a home-built microscope system. A circular mask and the objective's magnification control the diameter of the UV irradiation area.

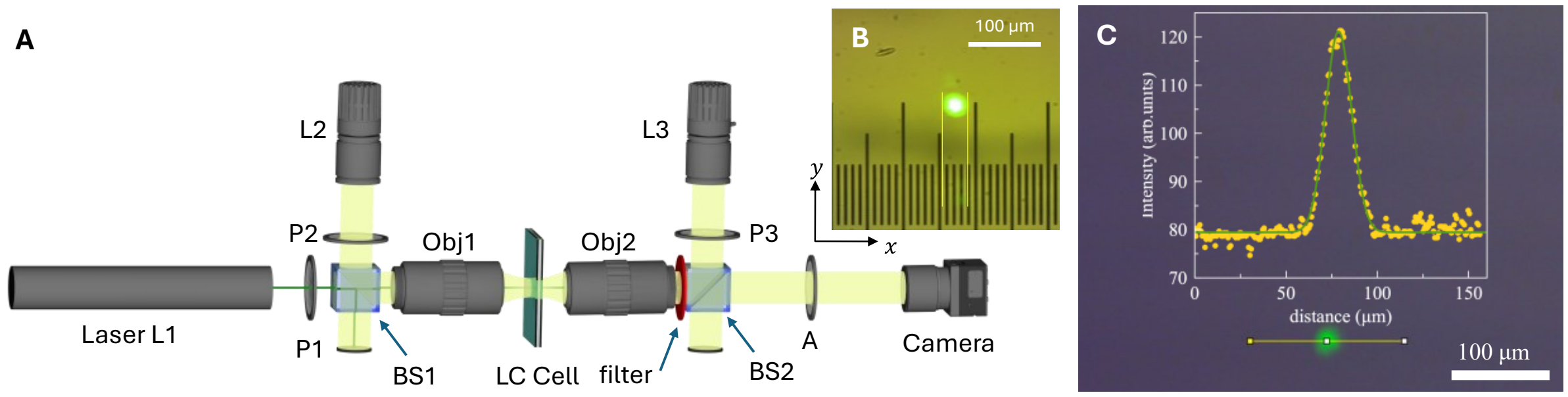

**Fig. 4. Experimental set-up.** (**A**) Linear polarizer P1 controls the polarization of a monochromatic (wavelength 457, 476, 488, 514, 532, and 1064 nm) light beam from the laser L1; two polychromatic light sources (L2, L3) with polarizers P2 and P3 record the textures in transmission and reflection modes, respectively; a 10X objective Obj1 reduces the laser beam diameter to ~36 µm. The laser power is measured after Obj1 using a laser power meter. 20X objective Obj2 images textures; a beamsplitter (BS1) couples the polychromatic L2 light and laser L1; a removable beamsplitter (BS2) is used only in reflection mode; a high-pass filter blocks the incident laser L1 beam; the analyzer A is placed before the camera to observe the textures of the cell as the function of the mutual orientation of P1 and A. The scheme was created using a Cheeta 3D software " https://www.cheetah3d.com/ with the help of Thorlabs components https://www.thorlabs.com. (**B**) The laser beam of a diameter 36 **µm** and wavelength 532 nm impinging on a cell. (**C**) The 532 nm beam profile is Gaussian.

Refractive indices measurements. The ordinary $n_o$ and extraordinary $n_e$ refractive indices for the N and $N_F$ phases of FNLC919 are determined using a wedge cell with an accuracy of 0.5%.

Polarizing optical microscopy observation of textures. The textures of the irradiated samples are observed in the assembled set-up with optical objectives and Thorlabs CMOS camera. PolScope MicroImager (Hinds Instruments) maps the cells' retardance and polarization **P** (which is along the optic axis of $N_F$). A polarizing optical microscope Olympus BX51 with an Amscope camera was used for observations of cell textures to verify the alignment and phases.

Measurements of photocurrent. The photocurrent is measured using a Tektronix TDS 2014 oscilloscope with a resistor connected in series to a planar cell with an electrode area of 1 $cm^2$, Figs.S5, S6. A DC bias voltage ranging from a $-5$ V to $+5$ V is applied using a Siglent SDG1032X waveform generator. UV light source of 365 nm (ThorLabs M365LP1) is used to illuminate the cell with intensities of 3 and 11 $mW/cm^2$.

Mathematica module. The module for solving Equations (4) and (6) and generating numerical data in Fig.3 is as follows:

```
fiNMBW[tTab_, eNorm_, wQ_, wP_, accuG_, shift_] :=
  Module[{tG, t, tC, tS, s, s0i, u, b, b0i, fB, fBi, fBiG, p, tpsuSol, fM, fMi,
    fMiG, fSol, k, deg, tpsuTab, dNorm}, deg = N[Degree]; dNorm = Sqrt[eNorm];
   fMi = tTab[[1]]/2 deg; fBi = tTab[[1]]/16 deg; tpsuTab = Table[ t = tG deg;
    fSol =
     FindRoot[
      {EllipticF[ArcSin[Sqrt[(2 (Cos[t - fM] - Cos[t - fB]))/((1 + Cos[t - fM]) (1 - Cos[t - fB]))]], (1 + Cos[t - fM]) / 2] ==
        dNorm, Sin[fB ]^2 (Cos[fB] wQ + wP)^2 == 4 eNorm (Cos[t - fM] - Cos[t - fB])},
      {{fM, fMi, 2 shift, t - shift}, {fB, fBi, shift, t - 2 shift}},
      AccuracyGoal -> accuG, PrecisionGoal -> accuG - 1]; fMi = fM /. fSol;
    fMiG = fMi / deg;  k = Sqrt[(1 + Cos[t - fMi]) / 2]; fBi = fB /. fSol; fBiG = fBi / deg;
    tpsuSol = Flatten[N[{tG, fMiG, k, fBiG, Sin[2 fBi ]^2}]]; tpsuSol, {tG, tTab}];
   tpsuTab];
```

**Acknowledgments**
The authors thank Merck Electronics KGaA, Darmstadt Germany for providing the material FNLC919 and Dr. Hari Krishna Bisoyi and Organic Synthesis Facility at the AMLCI for the synthesis and purification of DIO.

**Funding:**
National Science Foundation grant DMR-2341830 (ODL).
National Research Foundation of Ukraine grant 2025.07/0147 (VGN, OK, RK)
National Academy of Sciences project No. 0123U100832 (VGN, OK, RK)
Long-term program of support of the Ukrainian research teams at the Polish Academy of Sciences carried out in collaboration with the U.S. National Academy of Sciences with the financial support of external partners via the agreement No. PAN.BFB.S.BWZ.356.022.2023 (VGN, OK, RK, VS, MK).
NATO SPS project G6030 (VGN, ODL).

**Author contributions:**
Conceptualization: VGN, ODL
Methodology: SP, VGN, SS, ODL
Investigation: SP, OK, RK, MK, VS, BB, SS
Funding acquisition: VGN, ODL
Project administration: VGN, ODL
Supervision: VGN, ODL
Writing – original draft: SP, VGN, SS, ODL
Writing – review & editing: ODL

**Competing interests:** Authors declare that they have no competing interests.

**Data, code, and materials availability:** All data and code needed to evaluate and reproduce the results in the paper are present in the paper and/or the Supplementary Materials. This study did not generate new materials. The tabulated data for the plots in Fig.1D,E,F,M, Fig.3D, Fig.4C, Fig.S5B, and Fig.S6B are deposited in Materials Data Facility under the link https://www.materialsdatafacility.org/detail/347f924a-31ae-43d3-b8bf-c3550a12f0a1-1.0 with DOI: 10.18126/vpwd-n552

# Supplementary Materials for

## Realignment of ferroelectric nematic by photoinduced electric field

Sathyanarayana Paladugu et al.

*Corresponding author. Emails: vnazaren@iop.kiev.ua, olavrent@kent.edu

**This PDF file includes:**

Figs. S1 to S7

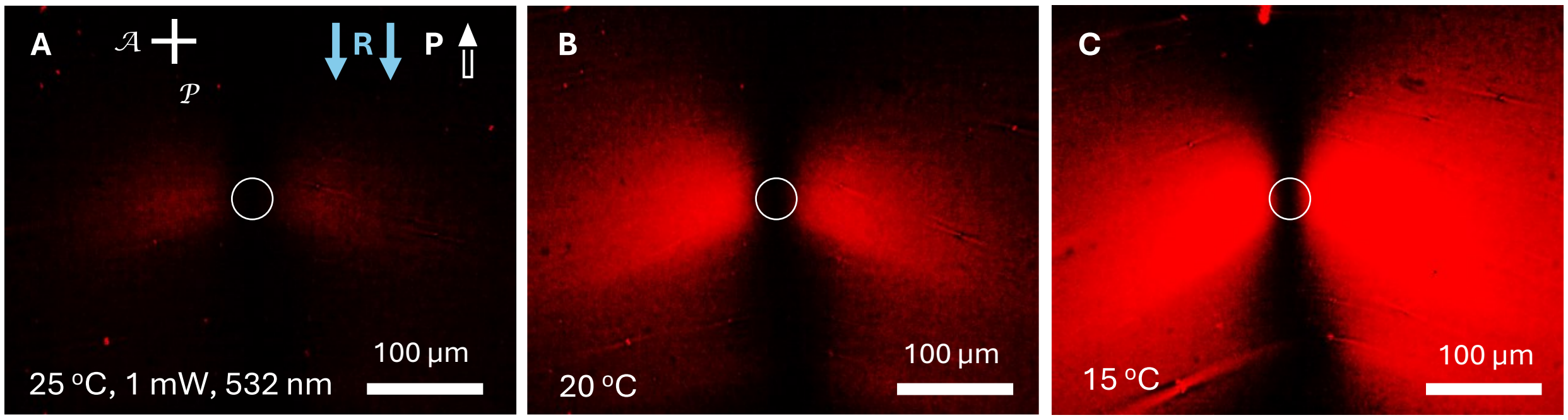

**Fig. S1. Temperature dependence of photo-induced reorientation of polarization at** (**A**) 25 °C, (**B**) 20 °C, (**C**) 15 °C. Reorientation increases with decreasing the temperature because of an increase of the polarization amplitude. Planar cell thickness 4.5 µm; λ = 532 nm; power 1 mW. Beam diameter 36 µm, alignment by two polyimide PI2555 rubbed coatings.

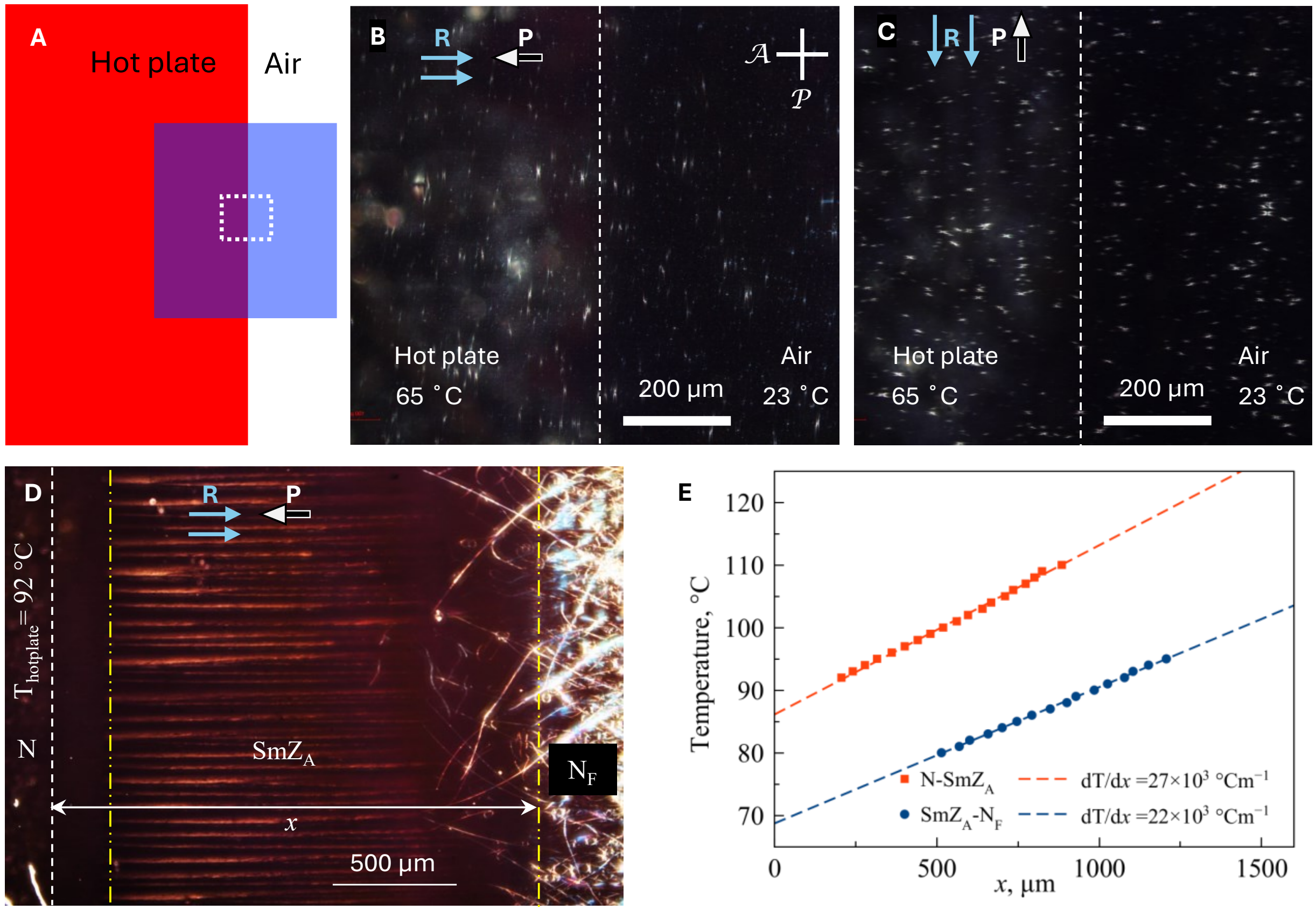

**Fig. S2. Polarizing optical microscopy textures of a DIO $N_F$ in a temperature gradient.** (**A**) Experimental setup; the dashed rectangle shows the area of observation. Cell is formed by two cover plates of thickness 150 µm. (**B**,**C**) The temperature gradient causes no realignment in the $N_F$ phase when the rubbing direction **R** and polarization **P** are parallel or perpendicular to the temperature gradient, respectively. (**D**) Coexistence of the N, $SmZ_A$, and $N_F$ phases across the temperature gradient. (**E**) Locations of the border lines between different phases used to calculate the temperature gradient. The cell thickness is 3.5 µm.

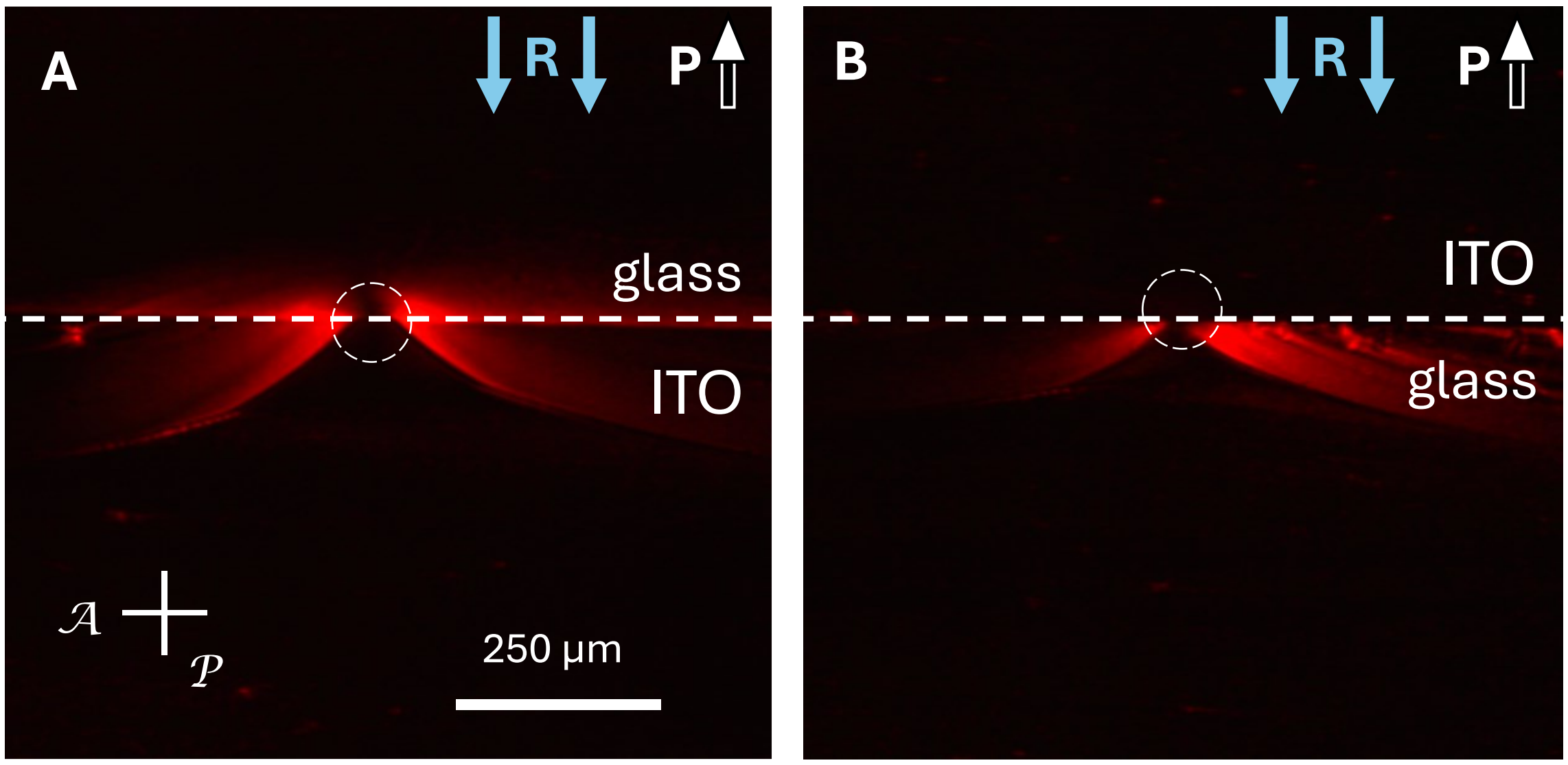

**Fig. S3. Polarization reorientation induced by irradiation of the FNLC919 $N_F$ at the ITO-glass interface.** (**A**) Reorientation occurs mostly in the ITO region. (**B**) Reorientation occurs mostly in the glass region. In both cases, reorientation is stronger in the regions that the polarization points away from, i.e., in the southern part of the pattern. $d$ = 4.5 μm, $\lambda$ = 532 nm, power 12.4 mW, T = 20 °C. Beam diameter ~100 μm.

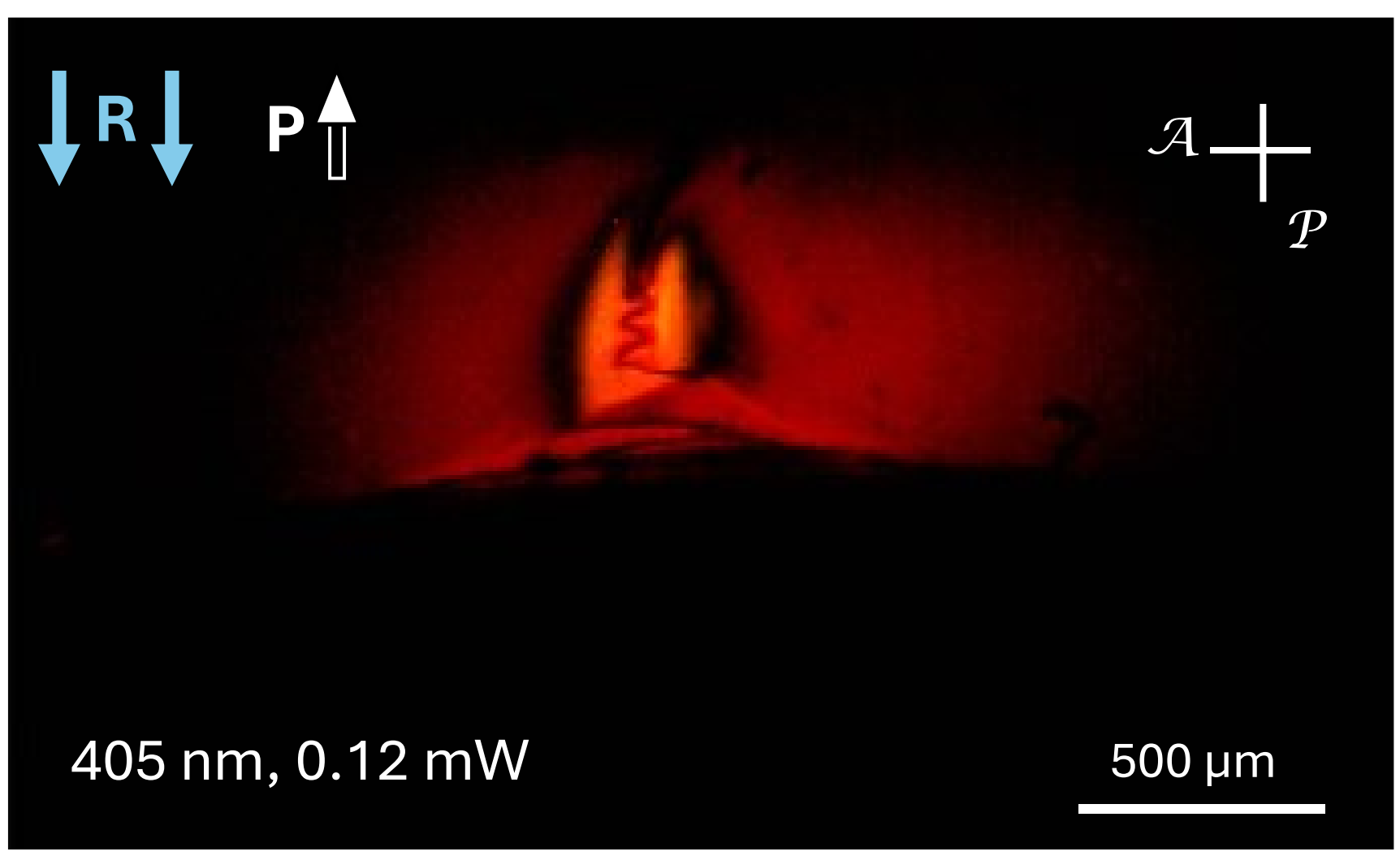

**Fig. S4. Polarization realignment by irradiation of a planar FNLC919 $N_F$ cell with two ITO electrodes and PI2555 alignment layers.** $d$ = 3.3 µm, $\lambda$ = 405 nm, power 0.12 mW, T = 20 °C. Beam diameter ~550 µm.

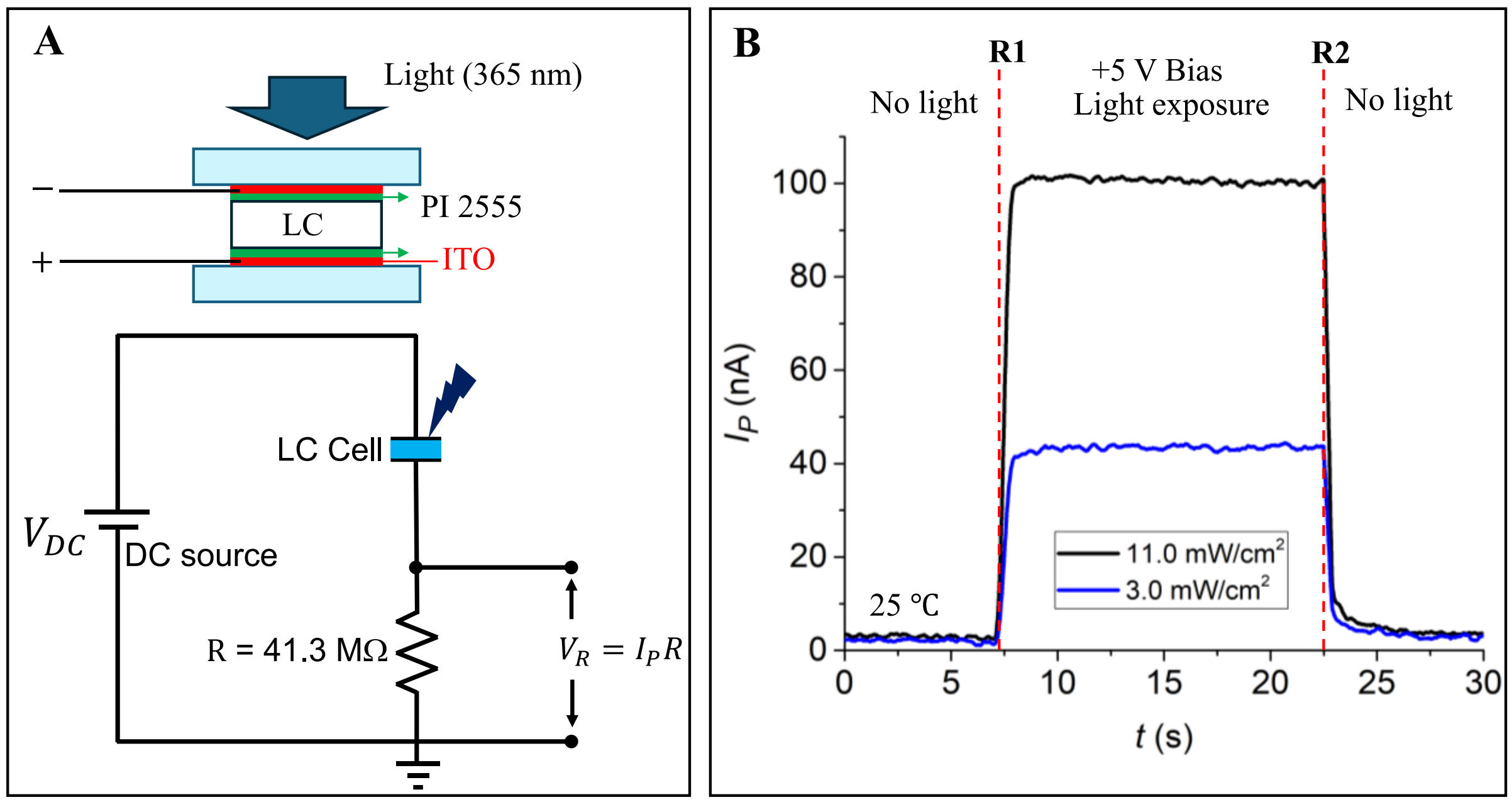

**Fig. S5. Photoinduced electric currents in a cell with ITO, polyimide PI2555, and N$_F$.** (**A**) The electrical circuit to measure the photocurrent in a planar N$_F$ cell caused by UV illumination at 365 nm. (**B**) Time-dependent photocurrents in a planar FNLC919 cell to a normally incident light beam with intensities of 3 and 11 mW/cm$^2$. The external DC bias voltage is $V_{\mathrm{DC}} = +5.0$ V. (R1) is the start of exposure, and (R2) is the end of exposure. $d = 3.8$ μm and $T = 25$ °C.

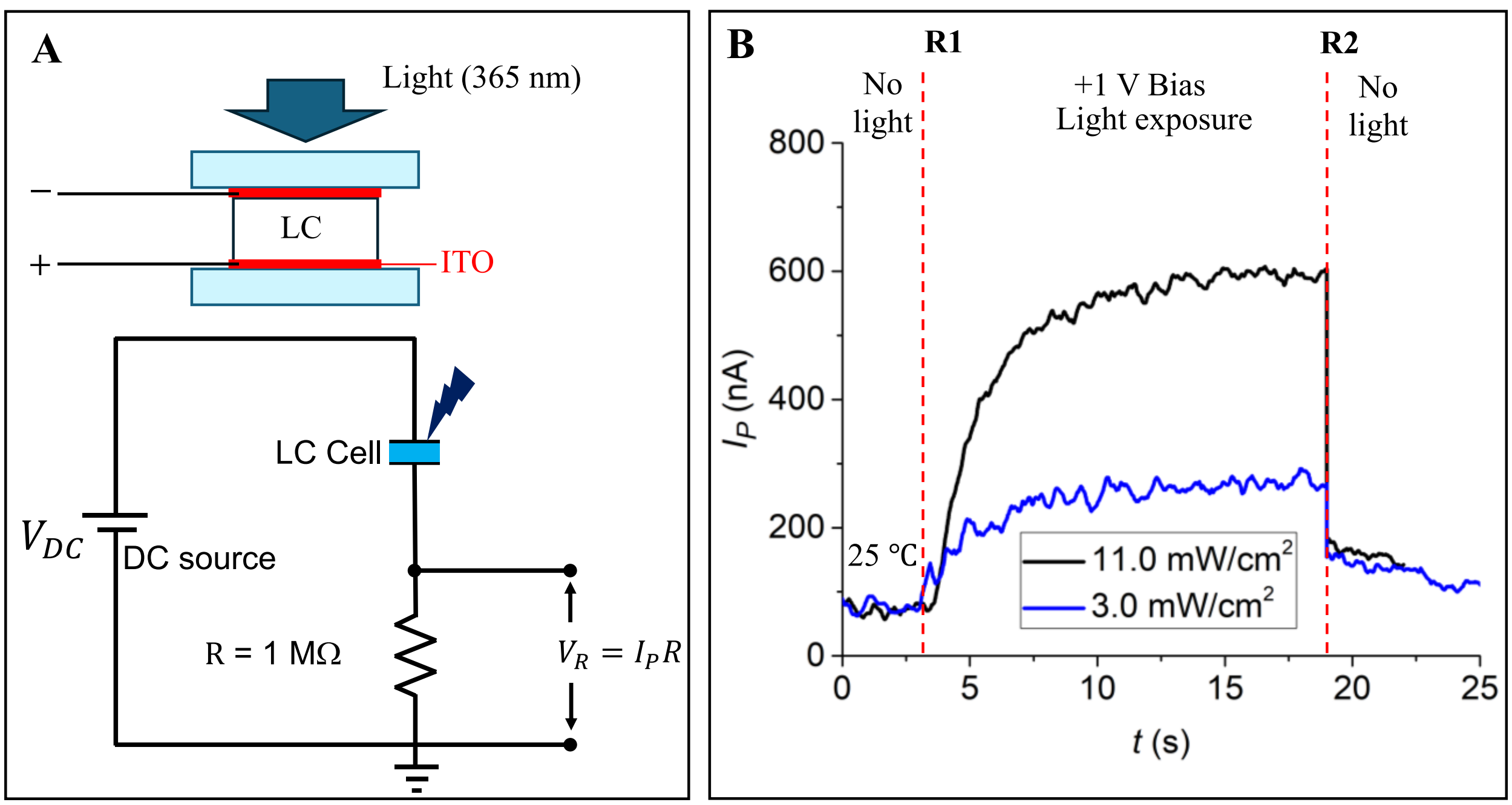

**Fig. S6. Photoinduced electric currents in a cell with ITO, N$_F$, but no polyimide PI2555.** (**A**) Schematic of the electrical circuit used to measure the photocurrent in a liquid crystal cell without an alignment layer upon illumination with 365 nm UV light. (**B**) Time-dependent photocurrents in a non-aligned liquid crystal cell (FNLC919; no rubbed polyimide layer) under normally incident light with intensities of 3 and 11 mW/cm$^2$. The external DC bias voltage is $V_{DC} = +1.0$ V. (R1) indicates the start of exposure, and (R2) indicates the end of exposure. $d = 1.7$ µm and T = 25 °C.

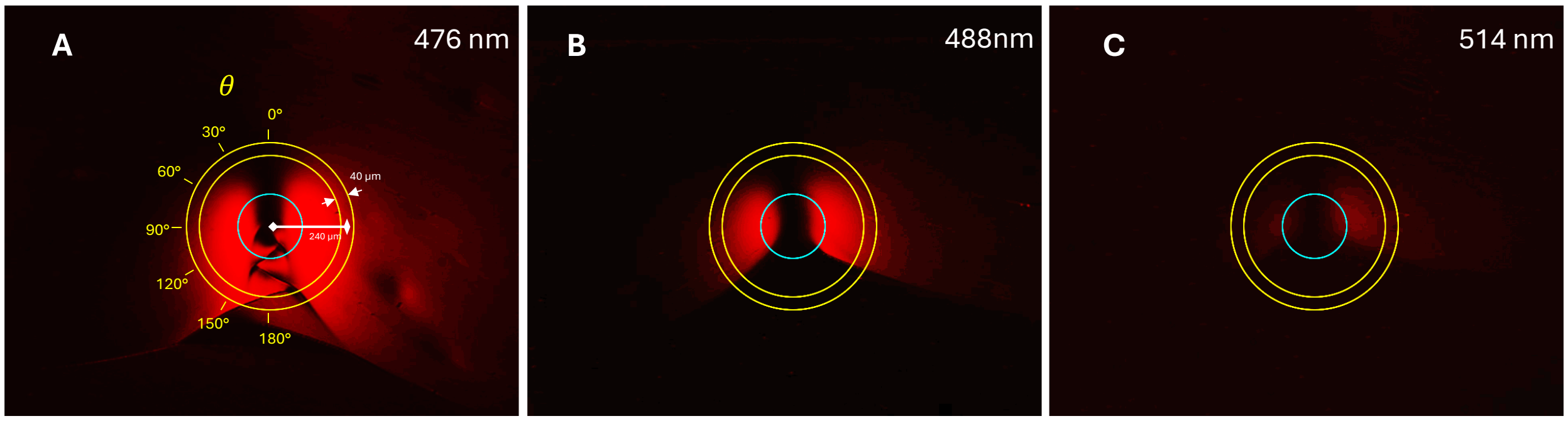

**Fig. S7. POM textures of polarization field in planar FNLC919 $N_F$ cell with two ITO electrodes and PI2555 alignment layers by irradiation at** (**A**) 476 nm, (**B**) 488 nm, and (**C**) 514 nm. The textures are used to calculate the azimuthal dependence of transmitted light intensity in Fig. 2. The transmitted intensity is calculated for an arch of a width 40 µm located at the distance 240 µm from the center of irradiated spot. The blue circle marks the waist of the light beam of a diameter 200 µm. $d = 4.5$ µm, power 5 mW, T = 20 °C.